\documentclass[twocolumn]{aastex63}
\usepackage{ragged2e}
\usepackage{natbib}

\newcommand{\curl}{ {\bf \nabla} \times}
\newcommand{\degree}{^\circ}

\def\p#1{{\color{magenta}{#1}}}

\def\j#1{{\color{purple}{#1}}}

\shorttitle{Magnetic Environment of a Stealth CME}
\shortauthors{O'Kane et al.}
\graphicspath{{./}{figures/}}
\begin{document}

\title{The Magnetic Environment of a Stealth Coronal Mass Ejection}

\author[0000-0002-8806-5591]{Jennifer O'Kane}
\affil{Mullard Space Science Laboratory, UCL, Holmbury St Mary, Dorking, Surrey, RH5 6NT, UK}

\author[0000-0003-1173-503X]{Cecilia Mac~Cormack}
\affil{Instituto de Astronomıa y Fısica del Espacio (IAFE), CONICET-UBA, Buenos Aires, Argentina}
\affil{Facultad de Ciencias Exactas y Naturales (FCEN), UBA, Buenos Aires, Argentina}

\author[0000-0001-9311-678X]{Cristina H.~Mandrini}
\affil{Instituto de Astronomıa y Fısica del Espacio (IAFE), CONICET-UBA, Buenos Aires, Argentina}
\affil{Facultad de Ciencias Exactas y Naturales (FCEN), UBA, Buenos Aires, Argentina}

\author[0000-0001-8215-6532]{Pascal D\'{e}moulin}
\affil{LESIA, Observatoire de Paris, Universit\'{e} PSL, CNRS, Sorbonne Universit\'{e}, Univ.Paris Diderot, Sorbonne Paris Cit\'{e}, 5 place Jules Janssen, 92195 Meudon, France}
\affil{Laboratoire Cogitamus, 1 3/4 rue Descartes, 75005 Paris, France}

\author[0000-0002-0053-4876]{Lucie M.~Green}
\affil{Mullard Space Science Laboratory, UCL, Holmbury St Mary, Dorking, Surrey, RH5 6NT, UK}

\author[0000-0003-3137-0277]{David M.~Long}
\affil{Mullard Space Science Laboratory, UCL, Holmbury St Mary, Dorking, Surrey, RH5 6NT, UK}

\author[0000-0001-7809-0067]{Gherardo Valori}
\affil{Mullard Space Science Laboratory, UCL, Holmbury St Mary, Dorking, Surrey, RH5 6NT, UK}

%\nocollaboration{6}

%% Note that the \and command from previous versions of AASTeX is now
%% depreciated in this version as it is no longer necessary. AASTeX 
%% automatically takes care of all commas and "and"s between authors names.

%% AASTeX 6.3 has the new \collaboration and \nocollaboration commands to
%% provide the collaboration status of a group of authors. These commands 
%% can be used either before or after the list of corresponding authors. The
%% argument for \collaboration is the collaboration identifier. Authors are
%% encouraged to surround collaboration identifiers with ()s. The 
%% \nocollaboration command takes no argument and exists to indicate that
%% the nearby authors are not part of surrounding collaborations.

%% Mark off the abstract in the ``abstract'' environment. 
\begin{abstract}

Interest in stealth coronal mass ejections (CMEs) is increasing due to their relatively high occurrence rate and space weather impact. However, typical CME signatures such as extreme-ultraviolet dimmings and post-eruptive arcades are hard to identify and require extensive image processing techniques. These weak observational signatures mean that little is currently understood about the physics of these events. We present an extensive study of the magnetic field configuration in which the stealth CME of 3 March 2011 occurred. Three distinct episodes of flare ribbon formation are observed in the stealth CME source active region (AR). Two occurred prior to the eruption and suggest the occurrence of magnetic reconnection that builds the structure which will become eruptive. The third occurs in a time close to the eruption of a cavity that is observed in STEREO-B 171\AA\ data; this subsequently becomes part of the propagating CME observed in coronagraph data. We use both local (Cartesian) and global (spherical) models of the coronal magnetic field, which are complemented and verified by the observational analysis. We find evidence of a coronal null point, with field lines computed from its neighbourhood connecting the stealth CME source region to two ARs in the northern hemisphere.  We conclude that reconnection at the null point aids the eruption of the stealth CME by removing field that acted to stabilise the pre-eruptive structure. This stealth CME, despite its weak signatures, has the main characteristics of other CMEs, and its eruption is driven by similar mechanisms.

\end{abstract}

%% Keywords should appear after the \end{abstract} command. 
%% See the online documentation for the full list of available subject
%% keywords and the rules for their use.

%% From the front matter, we move on to the body of the paper.
%% Sections are demarcated by \section and \subsection, respectively.
%% Observe the use of the LaTeX \label
%% command after the \subsection to give a symbolic KEY to the
%% subsection for cross-referencing in a \ref command.
%% You can use LaTeX's \ref and \label commands to keep track of
%% cross-references to sections, equations, tables, and figures.
%% That way, if you change the order of any elements, LaTeX will
%% automatically renumber them.
%%
%% We recommend that authors also use the natbib \citep
%% and \citet commands to identify citations.  The citations are
%% tied to the reference list via symbolic KEYs. The KEY corresponds
%% to the KEY in the \bibitem in the reference list below. 

\section{Introduction}

%INTRO TO STEALTH CMES 
Coronal mass ejections (CMEs) are large eruptions of solar plasma and magnetic field, expelled into the heliosphere at speeds ranging from a few tens to a few thousands of km~s$^{-1}$ \citep[see review by][]{webb2012coronal}. Stealth CMEs form a subset of all eruptive events and they are characterised by absent or faint signatures of eruption in the corona, with no obvious flaring, filament eruption, or strong EUV dimmings apparent. Following the first report of a stealth CME \citep{robbrecht2009no}, studies have shown common trends such as slow propagation speeds typically less than 500~km~s$^{-1}$ \citep{d2014observational}, a higher relative proportion of stealth CMEs at solar minimum \citep[$\sim$30\% of CMEs,][]{ma2010statistical} and an origin in the mid-corona from around 1.2 to 3.0 R$_\odot$ from Sun centre \citep{howard2013stealth}. The low speed and acceleration of mid-coronal stealth CMEs are then likely due to the low magnetic field strength and free magnetic energy present at those altitudes.
%\pc{I doubt that there is a limitation due to the Alfven velocity.  If forces are large enough, they could drive the ICME to a velocity larger than the Alfven velocity, as they involve gradient of the magnetic pressure (and tension).} \cmc{A small modification in the way I understand the sentence.}
In more recent works, image processing techniques have been able to enhance EUV and coronagraph data to reveal the fainter on-disc signatures associated with stealth CMEs \citep{alzate2017identification,o2019stealth}. These findings indicate that stealth CMEs often produce the same characteristic signatures as non-stealth CMEs, albeit weaker, meaning that the formation of the eruptive structure and the physical processes involved in stealth CMEs can be investigated from both observational and modelling perspectives.
%from both an observational and a modelling \p{with the same approach than for usual CMEs.} %perspective.
%\cmc{I went back to the original text. Pascal's modification can be implicit in the sentence but is not the same as was written before. Furthermore, the observational analysis requires improved techniques, they are not the same. Just see what you prefer.} 

CME eruption processes involve an energy storage phase, which may be the product of flux emergence and/or photospheric flows. Following this, an energy release phase sets in when ideal or non-ideal (resistive) processes lead to the rapid expulsion of the structure and the release of energy of the order 10$^{22}$-10$^{25}$J. The quasi-static evolution during the energy build-up phase may also be a period in which a magnetic flux rope is built via reconnection in the photosphere or chromosphere \citep{green2011photospheric} or in the mid-corona \citep{patsourakos2013direct,james2017disc}. The specific details of the pre-eruptive magnetic field configuration will then influence which mechanisms may act as the driver to produce the CME, e.g., the role of flare-related reconnection or an ideal magnetohydrodynamic instability \citep[for a review of CME processes see][]{green2018origin}. 

Stealth CMEs are expected to follow this energy storage and release sequence but the weak or absent signatures of flaring are an indication of only weak energy release associated with magnetic reconnection during the eruption itself. Kinematic studies suggest that the rise of a stealth CME follows an exponential profile, which is indicative of an instability \citep{o2019stealth}.

%Modelling of stealth CMEs
So far, few modelling studies have been conducted to shed light on the mechanisms behind stealth CME initiation. Comparisons to so-called streamer blowout CMEs have been drawn in which a streamer brightens and swells in the days prior to its eruption. Applying this scenario to modelling of stealth CMEs involves storing energy through slow shearing motions such as differential rotation, that displaces the footpoints of the coronal field along an extended polarity inversion line \citep{vourlidas2018streamer}, with reconnection playing the key role in ejecting the stealth CME structure and no pre-eruption flux rope necessarily being present \citep{lynch2016model}. The shortage of modelling studies is likely related to the relatively low number of detailed observational studies, caused by the difficulties of observing a relatively high-altitude structure formed in weaker magnetic field and within lower plasma density regions.

\begin{figure}[t!]
\centering
    \includegraphics[width=1\linewidth]{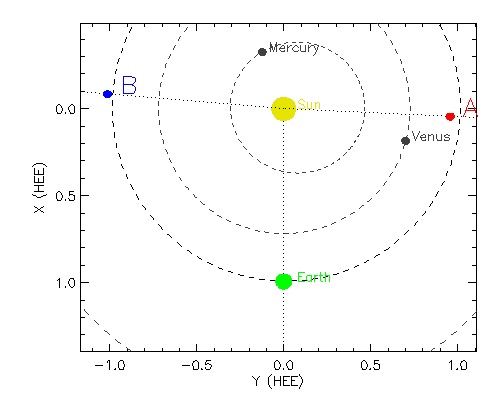}
    \caption{The positions of the STEREO spacecraft at 00:00 UT on 3-Mar-2011.}
    \label{fig:stpos}
  \end{figure}

%What we have done in this study 
This study bridges observations and magnetic field modelling for stealth CMEs. We build on the previous work of \citet{nitta2017earth} and \citet{o2019stealth} who found the source region of a stealth CME that occurred on 3 March 2011 to be NOAA active region (AR) 11165. In this work, we use local and global magnetic force-free field modelling to investigate the connectivity between the source region and its surroundings. Section \ref{sec:methods} outlines the data and methods used for this study. Section \ref{sec:event} summarises the event, the previous findings and the remote sensing results. Section \ref{sec:model} contains the modelling results. Section \ref{sec:disc} discusses all the final conclusions. 

\section{Data and Methods} \label{sec:methods}
   
\begin{figure*}[t!]
\centering
    \includegraphics[width=1\linewidth]{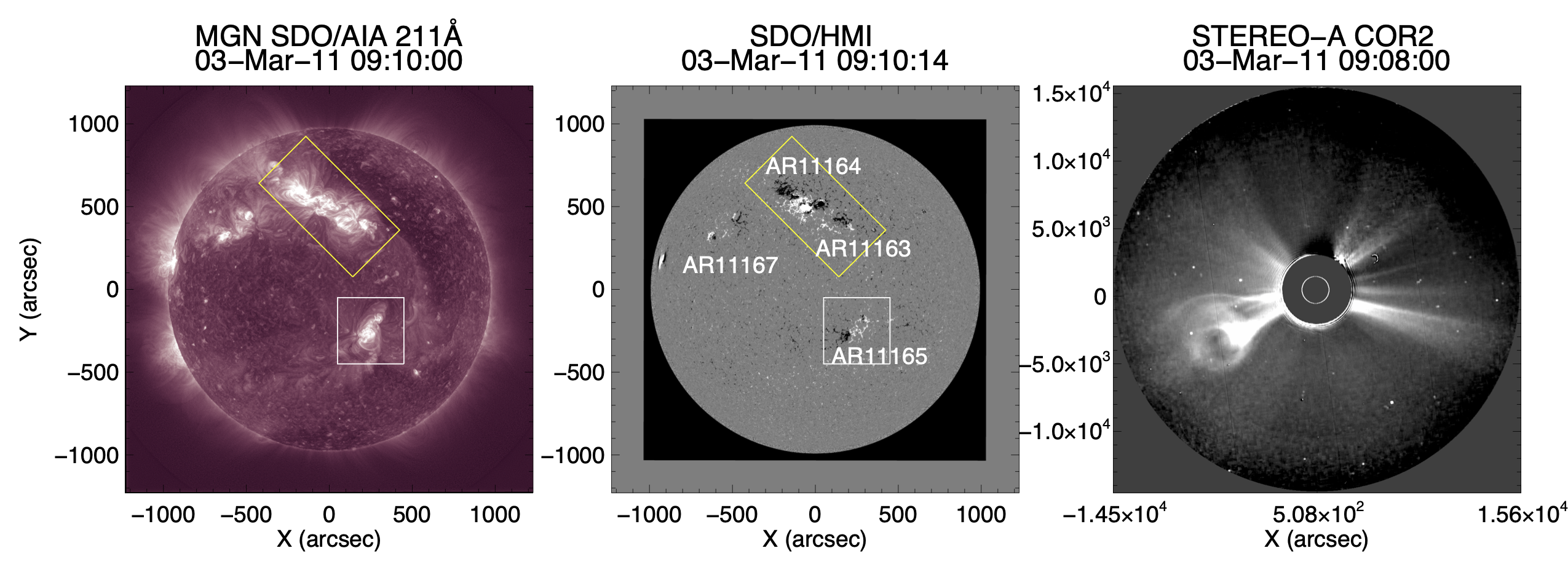}
    \caption{Left: MGN processed SDO AIA 211\AA\ image at 09:10 UT on 3 March 2011. The white box indicates the southern AR 11165, whilst the yellow box indicates the complex formed by ARs 11163 and 11164 in the northern hemisphere. Middle: The corresponding SDO HMI full-disc image. Right: STEREO-A COR2 image at 09:08 UT on 3 March 2011, showing that the stealth CME has a 3-part structure.
    %\pc{I would add labels of the 3 ARs on the magnetogram to facilitate text reading.} \cmc{I agree.}\jo{This has been done.}
    }
    \label{fig:cme}
  \end{figure*}

The evolution of the photospheric magnetic field in AR 11165 was analysed using the Helioseismic and Magnetic Imager data \citep[HMI,][]{scherrer2012helioseismic} on board the Solar Dynamics Observatory \citep[SDO,][]{pesnell2011solar}. HMI takes full disc images of the Sun, making narrow-band measurements around the photospheric line 6173~\AA\ and enabling the photospheric magnetic field to be derived. The AR flux is calculated using the Space Weather HMI Active Region Patches data series \citep[SHARP,][]{bobra2014helioseismic}. Only flux density values greater than  $\pm$150~G are considered. 

The evolution of the EUV corona in the time leading up to and during the stealth CME that was observed on 3 March 2011 is studied using data from SDO and the Solar Terrestrial Relations Observatory \citep[STEREO,][]{kaiser2008stereo}. Observations were obtained from the 304~\AA, 193~\AA\, and 211~\AA\ passbands by the SDO/Atmospheric Imaging Assembly \citep[AIA,][]{lemen2011atmospheric} with a 5 minute cadence. At the time of the eruption STEREO-A was 87$^{\circ}$ ahead of Earth and STEREO-B was 95$^{\circ}$ behind (see Figure~\ref{fig:stpos}). The 195~\AA\ (5 minute cadence) and 171~\AA\, (2 hour cadence) passbands from STEREO EUVI, which forms part of the Sun Earth Connection Coronal and Heliospheric Investigation instrument suite \citep[SECCHI,][]{howard2008sun}, were used to analyse the plasma emission structures of the stealth CME source region. The EUV data were subject to three image processing techniques.
\begin{enumerate}
\item The Multi-Scale Gaussian Normalization technique \citep[MGN,][]{morgan2014multi} that enhances small-scale structures in the corona and which has previously been used to identify signatures associated with stealth CMEs \citep{alzate2017identification,o2019stealth}. 
\item The Normalizing-Radial-Graded Filter \citep[NRGF,][]{morgan2006depiction} technique that enhances off-limb structures in EUV or white light coronagraph data.
\item Difference imaging that reveals dynamic changes in the corona. For stealth CMEs it is necessary to use temporal separations of 30 minutes or more \citep{nitta2017earth,o2019stealth} due to the relatively slow evolution of these events.
\end{enumerate}

The NRGF-processed COR1 and COR2 coronagraphs onboard the STEREO spacecraft were used to identify the stealth CME, and determine its plane-of-sky propagation direction and kinematics (mainly determined from the STEREO-B perspective). The combined EUV and coronagraph datasets provide overlapping fields of view; EUVI observes to a height of around 1.7 R$_\odot$ from Sun centre, COR1 observes from 1.5 - 4 R$_\odot$, and COR2 from 2.5 - 15 R$_\odot$. 

The kinematics of the eruption were determined using the COR1 data following the method of \citet{byrne2013improved}. This approach uses a residual re-sampling bootstrapping technique combined with the Savitsky-Golay algorithm to estimate the errors associated with the kinematics and derive the point-to-point velocity and acceleration of the CME. This approach has been shown by \citet{byrne2013improved} to be more rigorous than a normal numerical derivative approach and enables an estimation of the point-to-point kinematics that would not be possible using a simple fit to the distance-time data. 

\begin{figure}[t!]
\centering
    \includegraphics[width=1\linewidth]{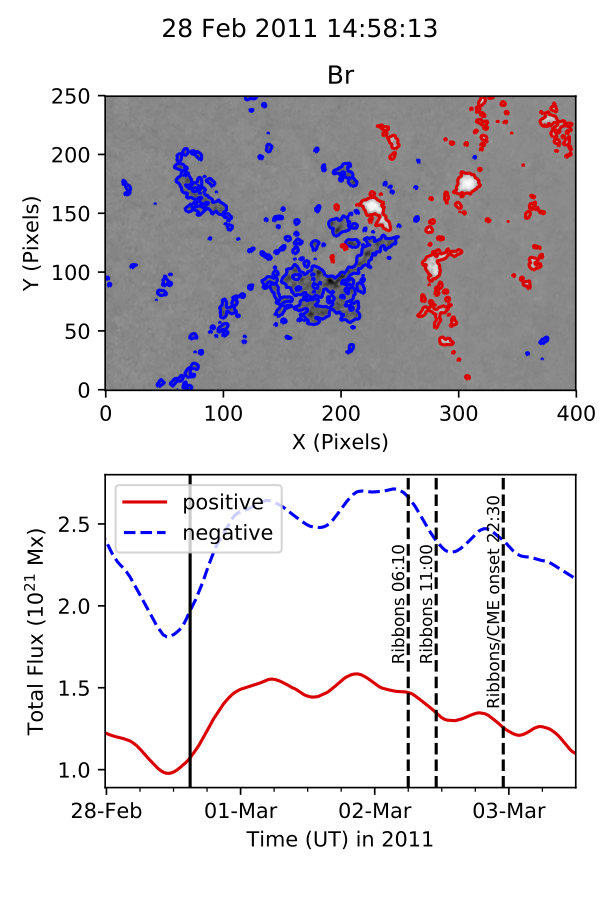}
    \caption{Radial magnetic field component of AR11165. The top panel shows a snapshot at 14:58 UT on 28 Feburary 2011, with negative and positive magnetic field surrounded by red and blue contours at $\pm 150$ G, respectively. The bottom panel shows the evolution of the positive and negative flux over the 84 hour period, before and after the stealth CME, from 00:00 UT on 28 February 2011 to 12:00 UT on 3 March 2011. The black solid line indicates the time of the HMI snapshot in the top panel, whilst the three dashed lines indicate the times of the three two-ribbon formation episodes discussed in Section \ref{sec:corona_evolution}. The third of these episodes is probably associated with the stealth CME initiation.}
    \label{fig:flux}
  \end{figure}

\section{Observational Overview} \label{sec:event}

The stealth CME was first detected in the Large Angle and Spectrometric Coronagraph (LASCO) C2 data as a partial halo at 05:48 UT on 3 March 2011. This event has previously been studied by \citet{nitta2017earth} who showed evidence of weak dimming signatures that indicated its source region was AR 11165. The analysis by \citet{o2019stealth} further supports this finding and additionally indicates that the CME could be the result of the eruption of a magnetic structure likely located at an altitude of $\sim$ 1.34 R$_\odot$ from Sun-centre, as determined from radio data. 
%seems a strong word since the structure is not seen, as far as I understand, you imply it exists because of the radio data; what about 
Figure \ref{fig:cme} shows the location of AR 11165 in the southern hemisphere near disc centre (left and central panels) and the clear %three-part 
circular cross-section structure of the ejecta as seen in STEREO-A COR2 data (right panel). In the northern hemisphere, a number of larger and more complex ARs were present. In this section we provide an overview of the evolution of the line-of-sight photospheric magnetic field, corona and the kinematics of the CME from an observational perspective.

\subsection{Photospheric Magnetic Field Evolution}
% Flux emergence at the internal PIL of active region 11165
AR 11165 has a small bipolar magnetic configuration. As detailed in \citet{o2019stealth} AR 11165 began to emerge on 25 February 2011 at the polarity inversion line of a previously decayed AR. In fact, the polarity inversion line was the site of repeated flux emergence with further episodes of emergence observed over a time period of $\sim$ 44 hours starting at 10:00 UT on 28 February 2011. Figure \ref{fig:flux} (bottom panel) shows the evolution of the radial component of the flux in the AR, with a 6-hour running average from 12 minute cadence data. The difference of approximately a factor of 2 between the negative and positive fluxes is likely due to the difficulty in distinguishing the negative flux emerging as part of AR 11165 from that of the pre-existing negative flux, some of which has been captured by the method used.

\subsection{Evolution of the corona prior to eruption}
\label{sec:corona_evolution}

\begin{figure*}[t!]
\centering
    \includegraphics[width=1\linewidth]{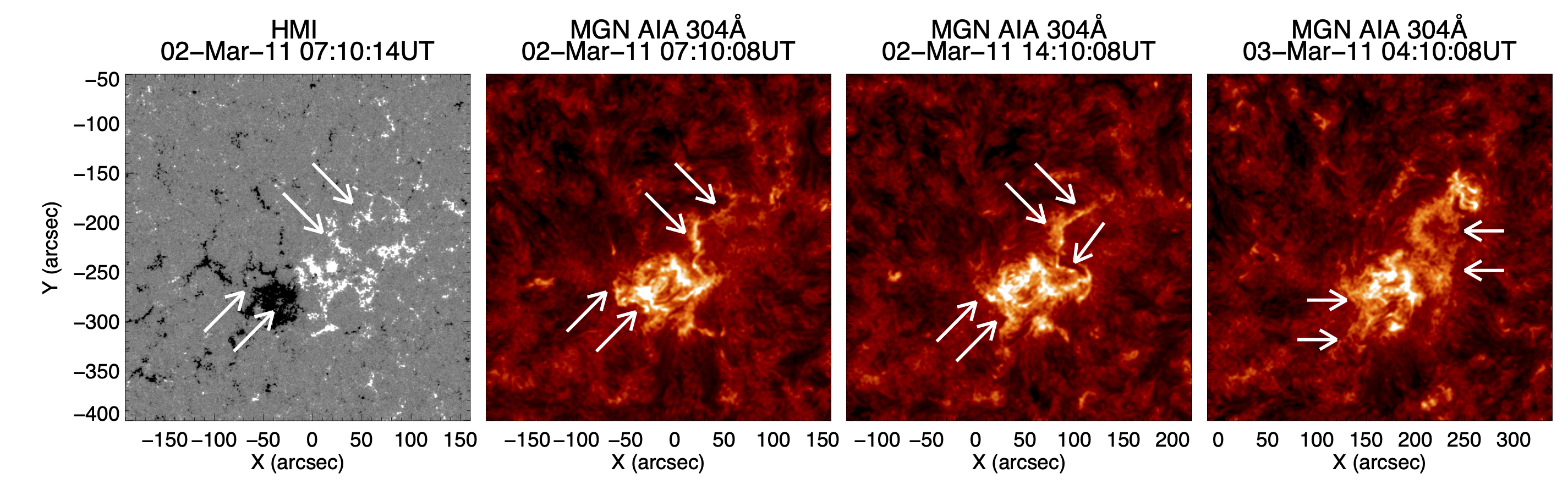}
    \caption{Two-ribbon formation as seen in the MGN processed AIA 304~\AA\,data (second, third and fourth panels) with HMI line-of-sight magnetogram for comparison (first panel). There are three main episodes of two-ribbon formation shown by the AIA 304~\AA\,data. The ribbon locations are indicated by the white arrows on the HMI magnetic map for the first episode.
    The first two episodes are associated with confined flaring and, as seen in the second and third panels, the ribbon locations are similar except for an elongation to the south of the northern ribbon (notice the additional arrow in the third panel). The third episode is probably associated with the eruption of the stealth CME; the ribbons move outwards and appear farther to the east and west of the AR center. In all cases the ribbons look patchy and their intensity is lower than that of the kernels located closer to the AR magnetic inversion line.
%\cmc{I commented Pascal's comment and modified the caption. I still think that the 304 MGN movie could really help to understand all this.}\jo{Have also added a sentence into the text clarifying that these are weaker than usually observed - but this is okay as it is a stealth cme!}
%    \pc{The central core of the AR looks brighter than the ribbons, OK? Could be worth to tell because I have difficulties to separate the ribbons from more persistent brightenings. Said differently, the ribbons are only the thin and not so bright features just next to the arrow heads, OK?  The peculiar distribution of the two ribbons is another difficulty to see them clearly. }
}
    \label{fig:ribbon}
  \end{figure*}
  
\begin{figure*}[t!]
\centering
    \includegraphics[width=1\linewidth]{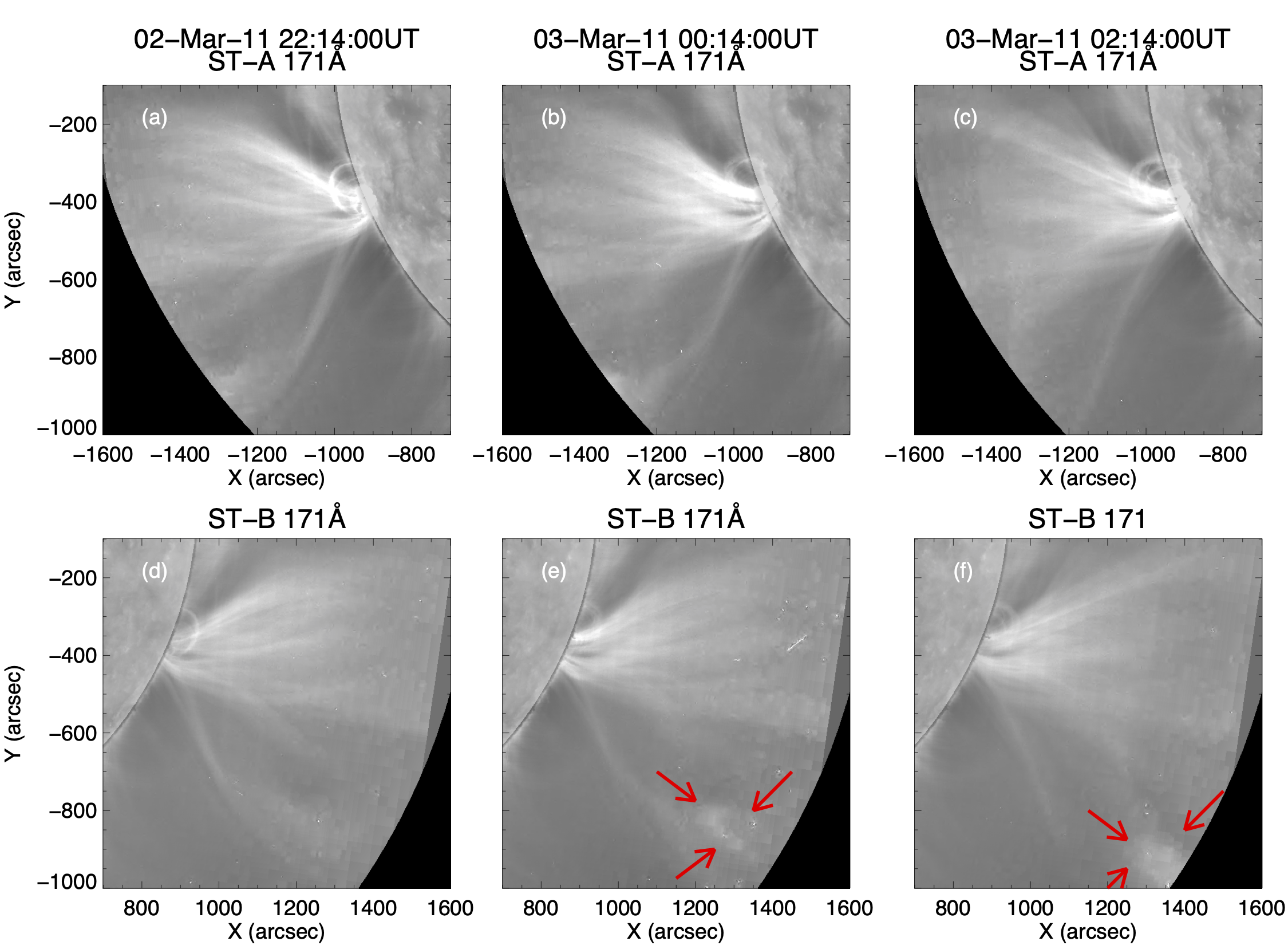}
    \caption{STEREO-A (panels a-c) and STEREO-B (panels d-f) EUVI 171~\AA\,images illustrating the configuration of AR 11165 and the larger-scale corona surrounding the active region. 
    EUVI-B data show an outward moving bright concave-up structure at 00:14 UT and 02:14 UT on 3 March 2011 as indicated by the red arrows in panels e and f. The concave-up structure is not seen in panel d.
}
    \label{fig:EUVI_cavity}
  \end{figure*}

% AIA 304 and the "flare" ribbons, AIA 193, 335 STEREO EUVI and the small flares
The MGN processed 304~\AA\ AIA data show that AR 11165 has a bright core interwoven with dark absorbing plasma threads (see Figure \ref{fig:ribbon}). The 304~\AA\ data also reveal three episodes of two-ribbon flare formation at the periphery of the AR bright core: two episodes occur after the flux emergence has ended (see Figure \ref{fig:flux}) but prior to the stealth CME and one occurs at the time of the stealth CME probable onset. The flare ribbons are faint and lack the classic hooks that would typically be observed in association with the eruption of a flux rope. However, given that stealth CMEs lack clear observational signatures due to their lower energy, weaker flare ribbons (or incomplete ribbons, because not all field lines undergoing reconnection have heated plasma at their footpoints) are to be expected. Then, the faint flare ribbons are in keeping with the stealth nature of the CME. Figure \ref{fig:ribbon} shows that the flare ribbons are at least partially rooted in the dispersed magnetic field of the previously decayed AR, rather than involving the newly emerging flux. This is most apparent in the flare ribbons that are located on the positive polarity field and indicates that the associated reconnection process involves structures that are overlying the emerging bipole. 
%\cmc{Jenny made an MGN movie of the evolution of AR 11165 in 304 and one can see what is described in this and the next three paragraphs. I have not found it attached to the previous paper. Is it too expensive or too much to attach it to this paper?}\jo{I was planning on attaching movies if it is okay to do so.}

The first episode of flare ribbon formation begins on 2 March 2011 at $\sim$06:10~UT (Figure \ref{fig:ribbon}, panel 2). The ribbons remain illuminated for around 4 hours and are also temporally coincident with a reorganisation of the corona as seen in AIA 193~\AA\ data. Details of the coronal evolution are given in \citep{o2019stealth} but in summary the lower altitude section of a loop structure apparently connected to the eastern ribbon dims and appears to expand at $\sim$06:10~UT on 2 March 2011. STEREO-A EUVI 195~\AA\ data also show the motion of this loop structure on the eastern side. From this perspective a southward motion of the loop leg is also detected.

The second episode of two-ribbon flare formation begins on 2 March 2011 at $\sim$11:00~UT and the ribbons remain bright for several hours fading by $\sim$15:00 UT on 2 March 2011 (Figure \ref{fig:ribbon}, panel 3). As is the case with the first flare ribbon episode, the ribbons of the second episode form and are extended on the western side further to the north than the AR main positive polarity concentration. Both the first and the second ribbon episodes are temporally coincident with coronal brightenings observed in the AIA 335~\AA\ waveband data at 06:30 UT and 15:45 UT on 2 March 2011. STEREO EUVI 195~\AA\ data also show a brightening of the active region and the formation of new loops at 07:30 UT and 11:30 UT, coincident in time with the first and second episode of ribbon formation respectively. However these brightenings are not sufficient to produce flares that are detectable in the GOES soft X-ray light curve. No separation of the two ribbons is observed in either episode in keeping with the confined nature of the flares (i.e. no eruptions were observed). This, along with the location of the ribbons at the AR periphery, indicates that the reconnection site is located above its core field. 

At $\sim$22:30~UT on 2 March 2011 the two ribbons activate for a third time, however this time they separate away from the AR core (Figure \ref{fig:ribbon}, panel 4). The two ribbons are observed to separate from each other over an approximately eight-hour period. Interpreting these flare ribbons in the context of the standard two-ribbon flare model and subsequent studies \citep[see][]{moore1980,kitahara1990,fletcher2004,qiu2009} would indicate that, during the first two episodes of two-ribbon formation, a reconfiguration of the coronal magnetic field above the newly emerged flux (AR core) occurs. This reconfiguration involves magnetic reconnection which most probably played a key role in forming the structure that then went on to erupt during the third episode of ribbon occurrence. This finding adds to the growing body of work showing that eruptive structures can form during reconnection episodes in the corona that produce confined flares and which could have a flux rope configuration \citep{patsourakos2013direct,james2017disc}. From the displacement of the flare ribbons with respect to a direction parallel to the polarity inversion line and the shear of the AR coronal loops we infer that  its magnetic field was of positive chirality, in agreement with the positive magnetic shear found in coronal loops (see also section \ref{sec:model-local} and Figure \ref{fig:lfff}).
 
%\pc{In sections 4 and 5, the trace of a flux rope is supposed to be present in observations. Is it so?  What are the evidences? The 2 ribbons?  Not obvious for me as the ribbons are patchy and not really typical of a flux rope (in particular I do not see hooks.} \cmc{Concerning this comment, there is a trace of the formation of a flux rope (twisted flux tube with an inverse-S shape, but no clear hooks) in the largest northern AR. This is seen in an MGN movie made by Jenny in 193 and also in 211 for the two northern ARs. I do not see a flux rope in AR 11165. However, we do not need to talk about flux ropes and we do not also need to stress the fact of the negative helicity in the northern ARs; therefore I have deleted the words flux rope and I have used erupting structure or similar.} \jo{Ok! I will find an appropriate location to just indicate that even though we don't see evidence of the flux rope itself - we so see observational support for reconnection, which would be observed during the building phase of a flux rope - but emission would be too weak for this event to observe the end product (if there is one - which we believe there is due to the reconnection in the lead up and the cavity observed in ST-B during the slow rise phase or the eruption.}

STEREO-A COR1 data show that prior to the stealth CME two other small and faint Earth-directed eruptions occur. The first being initially seen in the COR1 field-of-view at 06:00 UT and the second at 11:30 UT on 2 March 2011. However, these eruptions originate from the northern hemisphere and not from AR 11165. Their small and faint structure makes determining their specific source region challenging. A small eruption is observed from AR11167 which is likely the source of the CME observed at 06:00 UT. Meanwhile, AIA and EUVI data show ongoing dimming on the southern side of NOAA ARs 11163 and 11164 between 00:00 UT and 08:00 UT on 2 March 2011, which may be associated with the CME observed in COR1 at 11:30 UT. 

\subsection{Kinematics of the stealth CME eruption}

The lack of strong plasma emission from the erupting structure in EUV wavebands (and hence the classification of this event as a stealth CME) prevents a detailed analysis of the CME initiation and rise profile in EUV imaging data. However, a  bright concave-up structure, that is best seen in STEREO-B EUVI 171~\AA\ data, is observed at 00:14~UT (Figure~\ref{fig:EUVI_cavity}, panel e), 1.75 hours after the third episode of flare ribbon formation. The 2-hour cadence of EUVI 171~\AA\ data prevents a good analysis of the temporal evolution of this structure but it is seen again in the following image at 02:14~UT on 3 March 2011 (Figure~\ref{fig:EUVI_cavity}, panel f).  The concave-up structure observed in CMEs is interpreted as indicating the underside of a flux rope. 

\begin{figure}[t!]
\centering
    \includegraphics[width=1\linewidth]{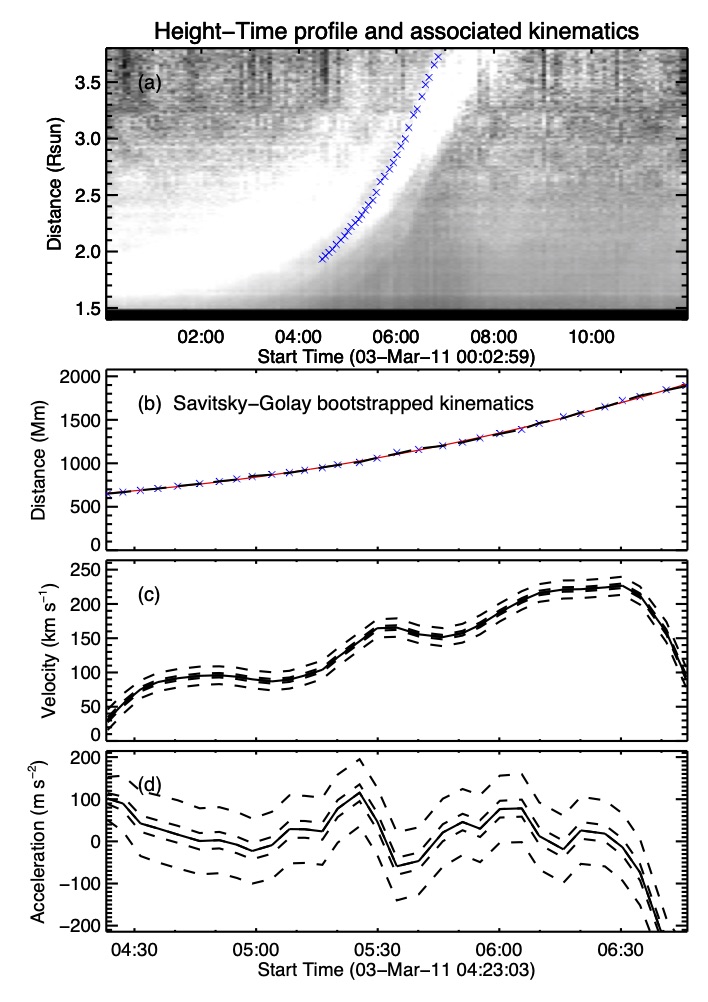}
    \caption{The height-time profile of the CME concave-up structure, with kinematics estimated using the Savitsky-Golay bootstrapping technique. Panel (a) displays the height-time image from the NRGF COR1-B data, with the manually tracked points as blue crosses.
    %\pc{worth to tell what is tracked (not the leading edge, as classical), especially because there is a bright diffuse region in front which is also going upward with a similar trajectory. So here a short summary, and a bit more in the main text.}  \cmc{It is said at the begnning of the caption. I would really prefer that the feature is shown in a top panel if possible, even if very difficult to see.}
    Panel (b) displays the tracked points as the median of the Savisky-Golay fit as black dashed lines. Panel (c) and panel (d) are the estimated velocity and acceleration profiles respectively, with the median (solid black line), interquartile boundaries (inner dashed lines), and upper and lower fences (outer dashed lines) all shown.
%\cmc{There is an error in the bottom labels of the panels or I am lost. Panel a says Start time (03-Mar-11 11:00:02.59) and the other panels say Start time (02-Mar-11 04:23:00); this has to be 03-Mar-11, OK?)}
%\pc{is the image not too saturated? I see a broad white corridor where the points cannot be located (but they were on the original data and their locations look precise, at odd with the background image).}\pc{they look very regularly located. Is the original image so sharp... or effect of your magic eyes?}\pc{The upper and lower fences look to define a too narrow interval on the right side where (unrealistic) fluctuations are larger} \cmc{What about cutting the time axis earlier to avoid some of the strong unrealistic fluctuations? I understand that then the top panel should finish earlier, but does it matter too much?}
%\jo{I have re-done the kinematics and improved the plot, let me know.}
%\pc{Much better !}
}
    \label{fig:speed}
  \end{figure}

Likewise, in the coronagraph data (COR1), there is no discernible structure at the leading edge of the stealth CME, only a concave-up structure at its trailing edge. The concave-up structure is followed in STEREO-B COR1 in order to determine the speed and acceleration of the eruption. However, it is not possible to confidently track the same plasma structure between EUVI data and coronagraph data due to the differing emission processes, so the kinematics of the CME are calculated only from STEREO-B COR data with a note that the EUVI-B concave-up structure is measured to be at a height of 1.54~R$_\odot$ from Sun-centre in the plane-of-the-sky at 00:14~UT and 1.67~R$_\odot$ at 02:14~UT on 3 March 2011 (Figure \ref{fig:EUVI_cavity} panels e and f), giving a very approximate plane-of-sky rise speed of 17 km s$^{-1}$. These heights are consistent with the underside of the stealth CME when it is first detected in STEREO-B COR1 data early on 3 March 2011 (see Figure \ref{fig:speed}). 
%\cmc{We are not showing this in the paper, OK? We are just saying it and showing nothing, it just sounds like "believe me". The same happens with the next paragraph.}

The top panel of Figure~\ref{fig:speed} shows the height-time variation of the erupting CME using data from STEREO-B COR1 along a radial slice at 120$^\circ$ measured in the clockwise direction from solar north. Figure~\ref{fig:speed} panel~b shows the identified evolution of the underside of the stealth CME fitted using a Savitsky-Golay bootstrapping technique \citep[cf.][]{byrne2013improved}. This approach enables the point-to-point variation of the speed and acceleration of the CME to be estimated as shown in panels c and d of Figure~\ref{fig:speed}. The initial measured speed of the white light CME, as determined from coronagraph data, at 04:23 UT on 3 March was found to be $\sim$ 31 km s$^{-1}$, with an acceleration of $\sim$ 102 m s$^{-2}$, consistent with the speed found from EUVI-B. From the kinematic data and the AIA 304~\AA\ flare ribbon formation we conclude that the stealth CME initiation began in the time period between 22:30 UT on 2 March 2011 (the time in which the third episode of flare ribbons begin) to 00:14 UT on 3 March 2011 (the time in which the concave-up structure is first observed in EUVI-B). 
%\cmc{The later time is just an upper boundary because of the concave-up structure you see in EUVI-B, OK?} 

\subsection{Transequatorial Loops}
\label{sec:transequatorial} 

\begin{figure*}[t!]
\centering
    \includegraphics[width=1\linewidth]{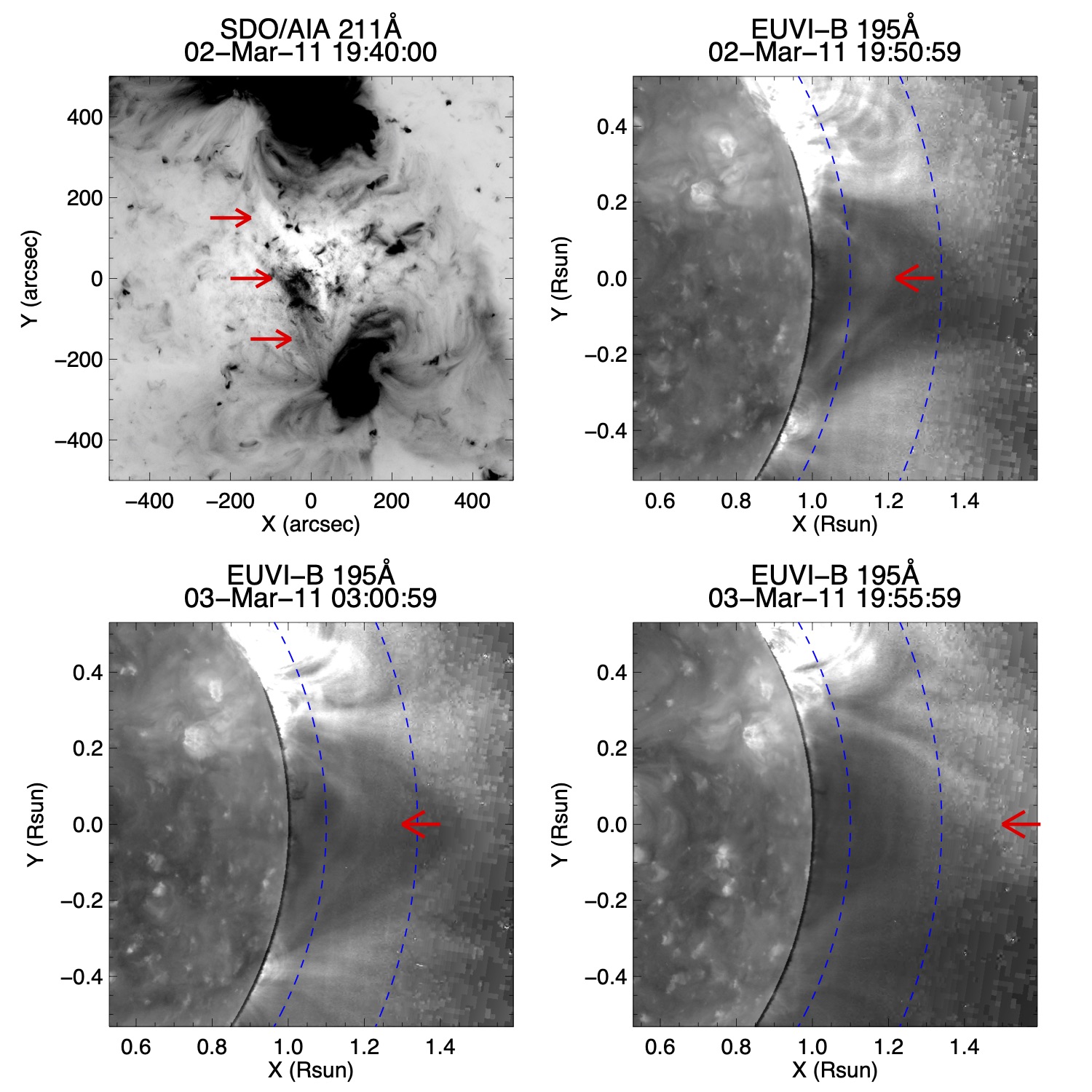}
    \caption{Top Left panel: NRGF-processed image of AIA 211~\AA\ showing the transequatorial loops 
    observed before the onset of the stealth CME, indicated by the red arrows. 
    Top Right panel: EUVI-B 195~\AA\ NRGF-processed image showing the transequatorial loops observed before the onset of the stealth CME, pointed out by the red arrow. 
    Bottom Left and Right panel: EUVI-B 195~\AA\ NRGF-processed image with the transequatorial loops observed after the onset of the stealth CME, pointed out by the red arrow.
    %\p{These loops are observed after the northern eruption.} \cmc{No Pascal, the two northern CMEs are on 2 March; these have to be the byproduct of the process that lead the stealth CME into the interplanetary medium. } 
    The blue-dashed lines indicate a height of 118~Mm and 234~Mm above the photosphere (1.1 and 1.34 solar radii from Sun center). 
    }
    \label{fig:tloop}
  \end{figure*}

%Transequatorial loops
The evolution of the corona at a larger scale, encompassing AR 11165 and the complex formed by ARs 11163 and 11164 to the north, is studied using SDO/AIA images and STEREO/EUVI data. As seen in STEREO EUVI 195~\AA\ waveband data from both Ahead and Behind spacecraft, transequatorial loops exist between these ARs in the northern and southern hemispheres.  STEREO-A shows that these loops are visible by 18:30~UT on 2 March 2011, while they are observed by 19:55~UT in STEREO-B. From the Earth perspective, the transequatorial loops are observed in AIA 193~\AA\ data around 19:40~UT on 2 March 2011 (Figure \ref{fig:tloop} left panel). These observations collectively show the presence of these large scale loops before the onset of the stealth CME but after the time period in which the coronal field above AR 11165 was reconfigured via magnetic reconnection as evidenced by the two-ribbon flares. We have searched for transequatorial loops in soft X-ray imaging data using Hinode/XRT; however, there is a gap in the full-disc XRT data between 2 March at $\approx$ 06:20~UT and 3 March 2011 at $\approx$ 06:20~UT.

% After Stealth CME eruption
After the stealth CME, the transequatorial loops are still observed, albeit at a higher altitude indicating that a reconfiguration occurred, perhaps due to reconnection driven by the expanding structure of the stealth CME. STEREO-A EUVI data indicate that these higher altitude loops can be seen by 08:25~UT on 3 March and in STEREO-B EUVI data by 19:55~UT on the same day. The loops are persistent and continue to be observed into 4 March 2011 at which time STEREO-B data show that the top of the transequatorial loops reach a height of  $\approx$234~Mm above the photosphere (1.34~R$_\odot$ from Sun-centre, see Figure \ref{fig:tloop}). The loops are not clear in AIA at this point, presumably due to the low plasma density of the loops against a higher density background.

\section{Magnetic Field Modelling}
\label{sec:model} 

To put the set of observations in a framework that allows us to understand the origin of the stealth CME, we have modelled the coronal magnetic field. 
We use two different approaches, a local model in Cartesian coordinates and two global models in spherical coordinates.

\subsection{AR 11165 Magnetic Field Model}
\label{sec:model-local} 

\begin{figure*}[t!]
\centering
    \includegraphics[width=1\linewidth]{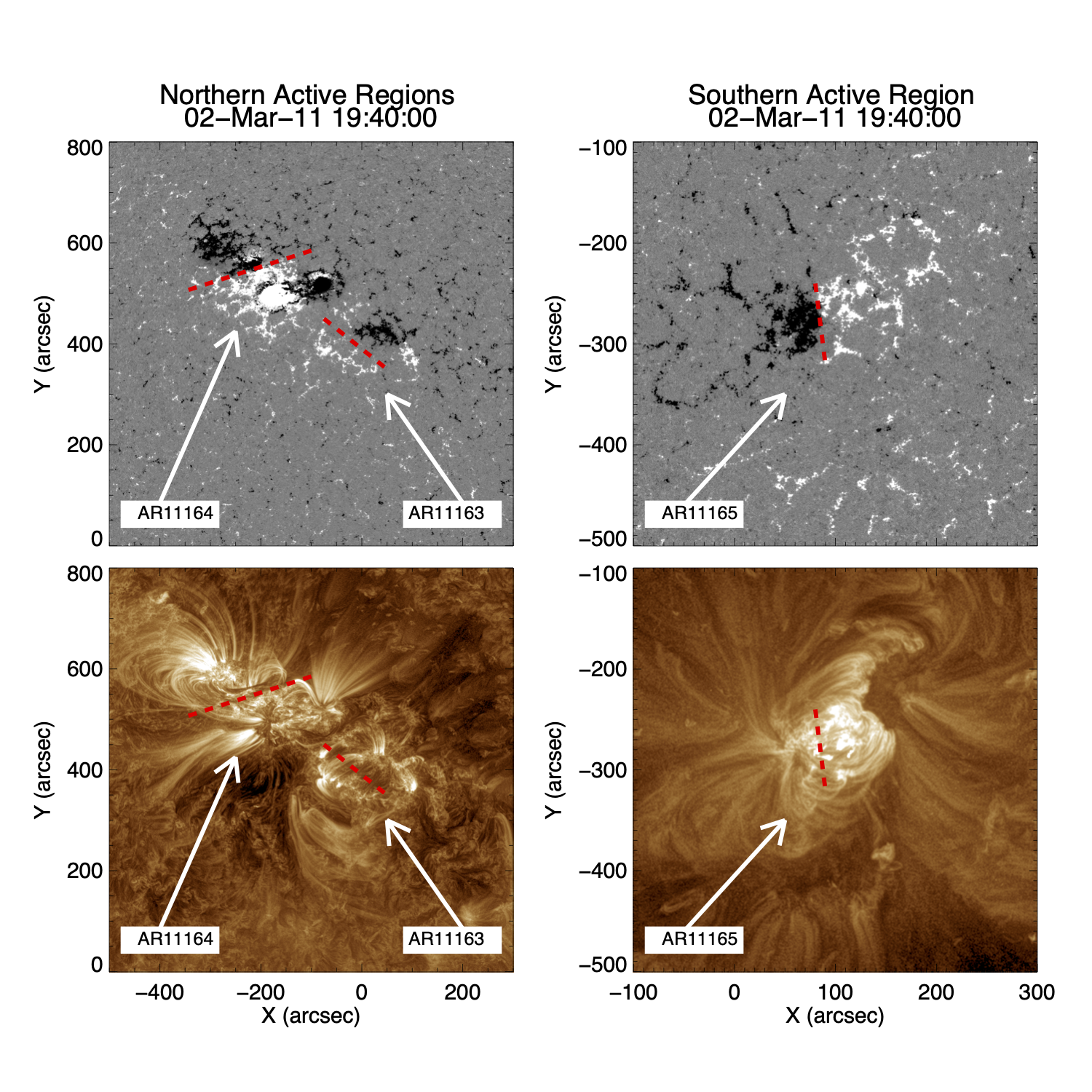}
    \caption{HMI (top two panels) and AIA 193~\AA\ (bottom two panels) images of the northern complex formed by AR 11163 and 11164 (left two panels) and the stealth CME source region, AR 11165 (right two panels).The main PILs for each AR are indicated by the red dashed lines. They have been drawn by eye approximating the main PILs with straight lines.  
%\cmc{ I propose to use the image at 19:40:07 for the northern ARs (it goes attached to my email), I think I see a dark inverse S-shape structure in the largest AR. I also propose not to include the magnetic field isocontours on the images. I believe it will be easier to see (at least) the global structure of the northern complex if we show the field in a different panel: the two magnetograms at the top and the two EUV images at the bottom. The rest of Pascal's remark is commented because I have changed and softened the text about the helicity sign.} 
%Looks to me weak to conclude on H sign as the loop shear is not large in the 3 ARs. Worth to strengthen as much as possible.and the vertical and horizontal axes show the position on the Sun in arcsec.\jo{Have changed the image to what I think you are after? Let me know if this is what you had in mind.}
}
    \label{fig:shear}
  \end{figure*}
  
\begin{figure}[t!]
\centering
    \includegraphics[width=1\linewidth]{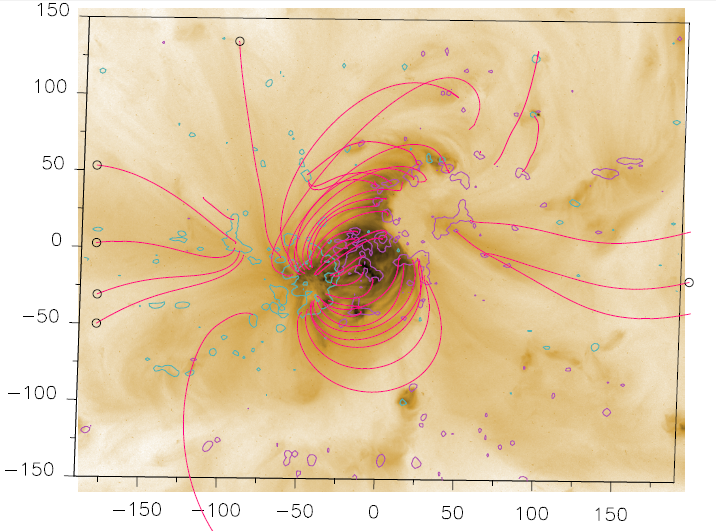}
    \caption{A LFFF model of AR 11165 for which we have used the line-of-sight HMI magnetogram shown in Figure~\ref{fig:shear} as boundary condition. Magnetic field contours at $\pm 50$, 200 G, positive (negative) shown in magenta (blue) color are overlaid on the AIA 193 \AA\ image closer in time to the magnetogram. A set of computed field lines, matching the global shape of the observed coronal loops has been added in continuous red lines. The circles present at the end of some field lines indicate that these lines touch one of the lateral or the top sides of the numerical box used for the drawing (not the one used to compute the field). The axes in this panel are in Mm and the figure is shown from the observer’s point of view.}
    \label{fig:lfff}
  \end{figure}

%\cmc{I have commented all the previous text and I am rewriting. The image in the paper for the northern ARs does not show a mixed helicity. I would say that the negative helicity is also marginal in that particular image. Therefore, I have softened all the text about helicity sign. As Pascal says, I also think the helicity sign is not relevant to this study; it was before, when we thought about including the interplanetary part. I have also not used the words flux rope.}  
%\pc{We do not really need the helicity sign for the following modelisation, and indeed all the paper.  I would write it short as a hint of a weak negative helicity for both ARs.  Telling more would need to compare with extrapolated field lines, or doing H flux maps at the photosphere... too side way for this paper.}\jo{Ok! Yes we had discussed softening the text on this as it is not too important for the main points in the paper!}
%\pc{I fully agree with this:} \cm{I do not know where this image shows a mixed helicity sign; at least it is not clear to me. I see that AR 11163 seems to have a global inverse S-shaped coronal structure and the inclination of the loops in AR 11164 seems to be also indicative of a negative helicity. So, I wrote what I see, but I kept what Jenny has written or may be I did not understand what she wrote.}
Figure~\ref{fig:shear}, two left panels, shows the photospheric magnetic field configuration and the coronal emission structures of the northern hemisphere active region complex formed by ARs 11163 and 11164. 
These ARs exhibit sheared loops as can be seen by the way they cross our approximated polarity inversion lines (PILs, depicted as red-dashed lines). In particular, the PIL in AR 11164 is quite complex and we only trace the main one. The global coronal structure of this AR displays roughly an inverse S-shape.   
The PIL of AR 11163 is much simpler and the loop inclinations to the PIL marginally also indicate a negative shear \citep[see e.g.][]{palmerio2017determining}.
 
Figure~\ref{fig:shear}, two right hand-side panels, show that AR 11165 has positively sheared loops as is expected from the hemispheric helicity trend for ARs in the southern hemisphere \citep{pevtsov2003helicity}. To confirm the sign of the shear, we have extrapolated the line-of-sight magnetic field of AR 11165 into the corona, using the HMI magnetogram under the linear force-free field (LFFF) approximation \citep[$\curl \vec B = \alpha \vec B$, with $\alpha$ constant,][]{Mandrini96,Demoulin97}. An example of the model is shown in Figure~\ref{fig:lfff}. The value of the free parameter of the model, $\alpha$, is set to best match the observed loops at the time of the magnetogram used for the extrapolation, following the procedure discussed by \citet{Green02}. The best-matching value is positive, $\alpha = 6.3 \times 10^{-3}$ Mm$^{-1}$. Our model also carries out a transformation of coordinates from the local AR frame to the observed one \citep[see][]{Demoulin97} so that our computed field lines can be compared to the AIA observed loops (as shown in Figure~\ref{fig:lfff}).

\subsection{The Three AR Complex Using a Cartesian Magnetic Field Model}
\label{sec:model-ARcomplex} 

Since the northern complex of ARs and the stealth CME source region are connected by transequatorial loops (see Section~\ref{sec:transequatorial}), we now investigate the magnetic connectivity between them. Because of the presence of moderate opposite magnetic shear in the three ARs, we extrapolate the same HMI photospheric magnetogram used to model the coronal field of AR 11165 (Figure~\ref{fig:lfff}) under a potential field approach, using an area that encompasses both the northern and southern regions as the boundary condition. A Cartesian model at such a large size-scale is disputable, however, its use allows us to consider a single magnetogram including the three ARs with a higher spatial resolution (i.e. our non-uniform grid size corresponds to HMI spatial resolution at the center of our computational box) and closer in time to the analysed events, reflecting more accurately the photospheric conditions. The results found in this section will be verified in Section~\ref{sec:model-global} using a global model in spherical coordinates.   

\begin{figure*}[t!]
\centering
    \includegraphics[width=0.8\linewidth]{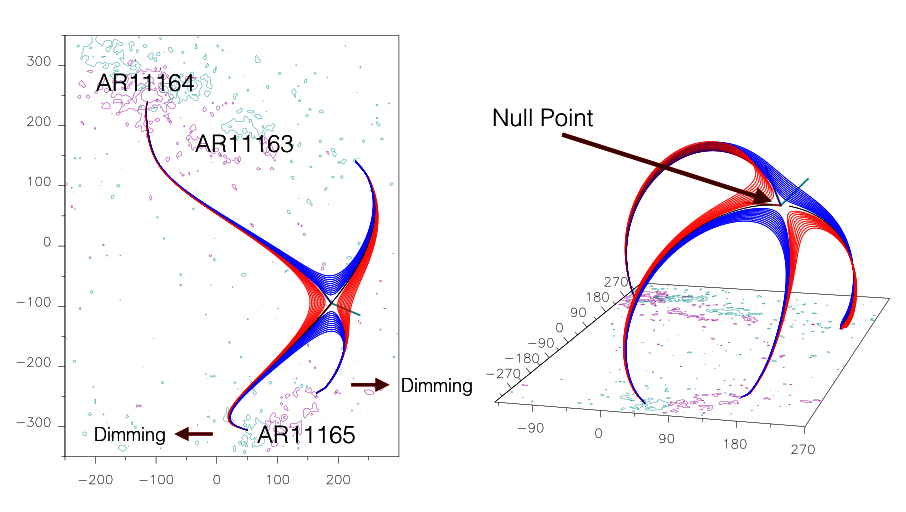}
    \caption{Coronal magnetic field model in the close vicinity of the magnetic null point found to the north-west of AR 11165. The left panel is drawn in the observer's point of view, while the right panel is a view from the solar south. 
    Field lines represent pre-reconnected (in blue colour) and post-reconnected ones (in red colour), as inferred from the observed evolution described in Sections~\ref{sec:corona_evolution} and~\ref{sec:transequatorial} and our interpretation in Section~\ref{sec:disc}. All axes are in Mm and the isocontours of the field correspond to $\pm 50, 200$ G in continuous magenta (blue) style for the positive (negative) values. The arrows point to the direction to which the dimming regions are observed to evolve, as shown in Figure 8 right panel in \citet{o2019stealth}.}
    \label{fig:pot_cartesian}
  \end{figure*}

We searched for topological structures in the extrapolated model and found the presence of a magnetic null point to the north-west of AR 11165 (see Figure~\ref{fig:pot_cartesian}). The local field connectivity around a magnetic null can be described using the linear term of the Taylor expansion of the field around such a point \citep[see][and references therein]{Demoulin94,Mandrini14,Mandrini15}. From the diagonalization of the Jacobian matrix of the field, we find three eigenvectors and the corresponding eigenvalues, which add up to zero to satisfy the field divergence-free condition. The eigenvalues are real for coronal conditions (a force-free field). A positive null point has two positive eigenvalues and conversely for a negative null. Figure~\ref{fig:pot_cartesian} illustrates the location of a negative  null point found at a height of 234 Mm above the photosphere (1.34 R$_\odot$ from Sun centre); this is indicated by the intersection of three segments that correspond to the directions of the three eigenvectors of the Jacobian matrix. These segments are colour coded to indicate the magnitude of the corresponding eigenvalues. For a negative null, dark blue (light blue) corresponds to the highest (lowest) negative eigenvalue in the null fan plane and red to the null spine eigenvalue. 
%\cmc{The height of the null point in this model is the same as the height of the radio source found in Jenny's previous paper, has this some meaning? Should we stress this or not because the height in the global models are lower?}\jo{I do not think it needs to be stressed too much, but it is worth noting? Opinions?} The radio source is to the south whereas the null to the north, so I would not link them together. 

We trace sets of field lines in the neighbourhood of the null point, with starting positions in the direction of the three eigenvectors, to explore the different connectivity domains. This connectivity is illustrated in both panels of Figure~\ref{fig:pot_cartesian}. The left panel corresponds to the observer's point of view along the Sun-Earth line, while the right panel shows the location of the null point as viewed from the solar south. The field lines in both panels have been drawn in blue and red to indicate what we envisioned could be the pre-reconnection set (blue field lines) and the post-reconnection set (red field lines), following the evolution of the emission structures and transequatorial loops discussed in Sections~\ref{sec:corona_evolution} and~\ref{sec:transequatorial}. The illustrated connectivity shows that the positive polarity of AR 11165 is connected to the dispersed negative  polarity of AR 11163 and its negative polarity to the positive polarity of AR 11164. The footpoints of the field lines in AR 11165 are located 
%\cm{within the} \cmc{Jenny, please check, but when I compare the footpoints of my extrapolated field lines and your dimming evolution I see the footpoints within the dimming regions.} Vicinity is a bit more vague to avoid any confusion or extra proofs within the paper.
in the vicinity of the twin EUV dimmings observed in association with the stealth CME, as shown in Figure 8 of \citet{o2019stealth}. The location of the dimming regions, their evolution over a 9 hour period and their relationship to the CMEs are discussed in detail in that article. 
%In our present analysis we add the close spatial relationship and possible interaction of the field lines in the neighbourhood of the null point and the magnetic field that erupts as the stealth CME.

 \begin{figure*}[t!]
\centering
    \includegraphics[width=0.9\linewidth]{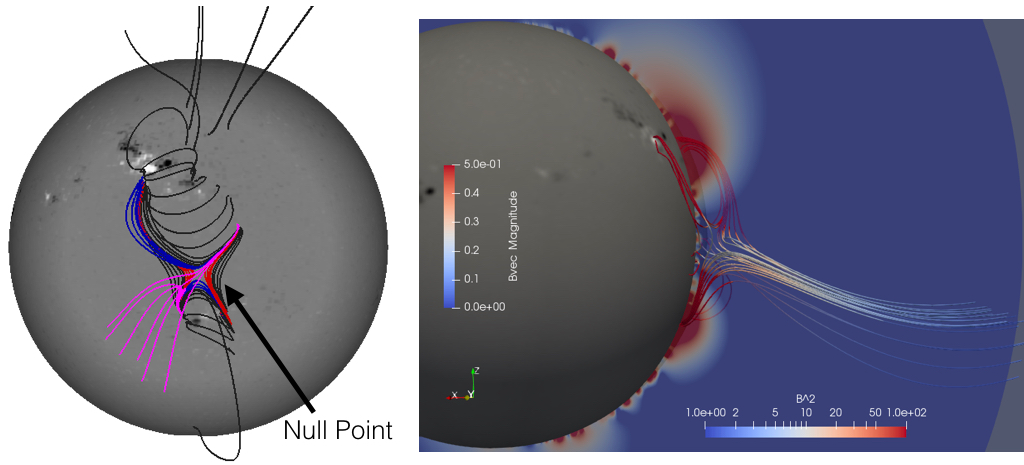}
    \caption{PFSS model of CR 2107. In the left panel, the magnetic field is computed with the FDIPS code. The null point is viewed when AR 11165 is numerically rotated to the central meridian.
    A set of field lines has been computed from its neighbourhood using the same blue and red convention as in Figure~\ref{fig:pot_cartesian}.  These field lines show a similar connectivity as the ones in the Cartesian field model. Closed field lines are added in black mostly for context, while open ones are shown in pink. 
    In the right panel, the model is computed using the pfssy code.  We illustrate the connectivity surrounding the null point setting AR 11165 on the west limb. This shows the shape and direction of the `open' field. The vertical coloured bar represents the magnitude of the field in Gauss along the field lines. The Python code also includes the computation of $B^2$, within the slice passing through the centre of the null point (bottom right color scale). 
    }
    \label{fig:pfss}
  \end{figure*}
  
\subsection{Global Magnetic Field Model}
\label{sec:model-global}

The magnetic null point found in the local Cartesian model could play a crucial role in facilitating the initiation of the stealth CME; therefore, we have searched for its presence in a global potential field source surface (PFSS) model, which could also give us clues about the location of `open' field lines, their shape, and their possibility of channeling the stealth CME and influencing its propagation direction into the interplanetary (IP) medium. Because of the relevance of the role of this null point, we have checked the robustness of its presence using two different approaches to compute the PFSS global model. 

Both PFSS models use as their lower boundary condition the HMI magnetic field synoptic map from Carrington rotation (CR) 2107, that ran from 16 February to 16 March 2011. HMI synoptic maps are computed from line-of-sight magnetograms by combining central meridian data from 20 magnetograms collected over a 4-hour interval. A synoptic map is made with the magnetograms collected over a full solar rotation ($\approx$ 27.7-day interval) with 3600 $\times$ 1440 steps in longitude and sine latitude.  Both models assume that the magnetic field becomes purely radial at a height, called the source surface, which is set to the value 2.5 R$_\odot$.  Full details concerning the construction of synoptic maps can be found in the HMI web-site (http://jsoc.stanford.edu/jsocwiki/SynopticMaps).

Our first PFSS modeling approach uses the Finite Difference Iterative Potential-Field Solver (FDIPS) code described by \citet{Toth11}. The FDIPS code, which is freely available from the Center for Space Environment Modeling (CSEM) at the University of Michigan (http://csem.engin.umich.edu/tools/FDIPS), makes use of an iterative finite-difference method to solve the Laplace equation for the magnetic field. The spatial resolution of this particular model is 1$\degree$ in longitude (360 longitudinal grid points), 0.11 in the sine of latitude (180 latitudinal grid points) and 0.01 R$_\odot$ in the radial direction. We searched for a null point using a method similar to that discussed in the previous section but using a spherical geometry for the coronal field. Figure~\ref{fig:pfss} left panel shows the location of a null point and the connectivity of the field when the Sun is rotated so that AR 11165 is facing the observer. This null point is located at a height of $\approx$ 118 Mm, lower than in the Cartesian model, but the connectivity clearly resembles the one of that model (Figure~\ref{fig:pot_cartesian}). 

Our second approach uses the PFSS solver in Python \citep[pfsspy,][]{https://doi.org/10.5281/zenodo.1472183,pfss_stansby_2019}. In this case we re-sampled the original field spatial resolution to a 540 and 270 pixels in longitude and latitude, respectively. 
%\pc{I try to solve the LaTeX problem with pfss\_yeates\_2018 reference but I do not succeed. Is this Software command fine? I never used it.  Also the name sample63.tex is not really appropriate!}
This solver also uses finite differences, and follows the method in \cite{Ballegooijen2000} to effectively compute discrete spherical harmonics global functions. We also found a magnetic null point at a similar location and with a similar connectivity in its neighbourhood as before, but at a height of $\approx$ 104 Mm. Therefore, though there are differences that come from the different approaches used, and different boundary conditions between the local and global models, the presence of this null point between the AR complex in the north and the stealth CME source region is consistent. Finally, both PFSS models suggest that there are `open' field lines above the stealth CME source region that are inclined to the south (Figure~\ref{fig:pfss}, mainly the right panel). 

The inclination of the `open' field lines as well as the reaction of the surrounding potential field, present over AR 11165, may influence the CME propagation direction \citep[e.g.,][]{cremades2006properties,gopalswamy2009cmeCH}. The erupting magnetic field is both compressing and bending the surrounding magnetic field.  This induces a reactive force with both a gradient of magnetic pressure and a magnetic tension, with both forces proportional to $B{^2}$.
$B{^2}$ is seen to be higher above the two northern hemisphere ARs, lower in the region of the null point, and increases again over AR 11165 (see Figure~\ref{fig:pfss}). 
The CME is therefore likely to be influenced by $B{^2}$ present in the surrounding potential field, in agreement with previous studies  \citep[][and references therein]{gui2011cmeDeflection, kay2013ForeCAT, sieyra2020analysis} and be deflected southward contrary to the general equatorial deflection of many CMEs \citep{Fwarcremades2004three}. The presence of open field lines, deflected towards the south, is also expected to deflect the CME along them \citep[][and references therein]{makela2013CHinfluence, wang2020deflection}.

\begin{figure}[t!]
\centering
    \includegraphics[width=1\linewidth]{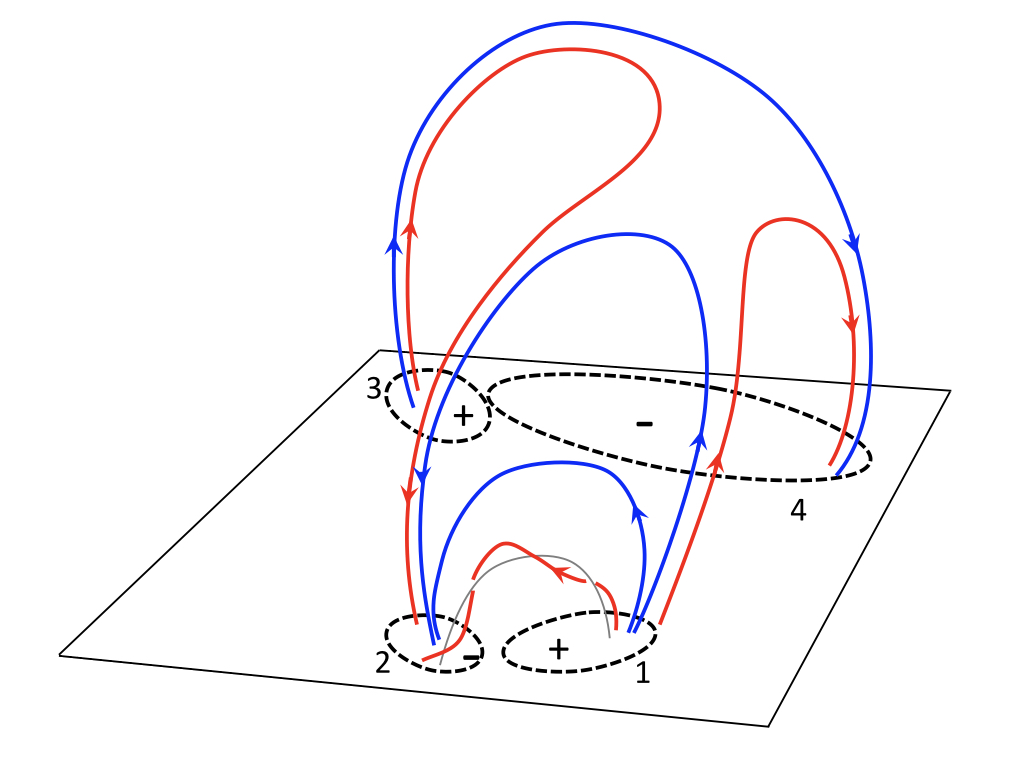}
    \caption{%have modified Pascal's changes to clarify 
    The sketch illustrates the scenario we propose for the origin of the stealth CME in Section~\ref{sec:disc}. The erupting structure is drawn in red within AR 11165 (with footpoints in magnetic polarities 1 and 2). This structure is probably the result of the two episodes of two-ribbon formation occurring before the stealth CME. The blue field lines located above the erupting structure belong to the arcade which will reconnect at the null point with the set of blue field lines (for which we only show one) associated to the AR complex in the north (simplified to only include the relevant magnetic polarities as 3 and 4).  This reconnection process will result in the set of two red field lines representing the transequatorial loops discussed in Section~\ref{sec:transequatorial}.}
    \label{fig:sketch}
  \end{figure}

\section{Conclusions} \label{sec:disc}

This study analyses the stealth CME of 3 March 2011 and seeks to gain insight into the magnetic environment in which the eruption occurred. Detailed data analysis and the use of image processing techniques have revealed observational signatures of the evolution of the magnetic field toward eruption that are then interpreted in the framework of three potential magnetic field coronal models.  

In summary, flux emergence dominates AR11165 in the lead up to the eruption, which appears to play a key role in creating the conditions sufficient for magnetic reconnection in the pre-existing arcade. Episodes of reconnection are evidenced by the observation of two-ribbon flares in the stealth CME source region (see Section~\ref{sec:corona_evolution}). Two episodes of magnetic reconnection occur in the decayed sheared field above the emerging bipole of AR 11165 forming the structure that later erupts as the stealth CME. All coronal models show a null point whose associated field lines connect AR11165 in the southern hemisphere and ARs 11163 and 11164 in the north. The presence of a coronal null point (and reconnection at this null) in the large scale field above the upper arcade of the pre-eruptive structure, which may be a flux rope, means that a rapidly decreasing field strength with height could enable the rope to become unstable in the hours after it is formed. Upon eruption, the stealth CME is deflected to the south, due to the ‘open’ field, and the overall variation in magnetic pressure. The eruption drives further reconnection at the null point creating new transequatorial loops that are observed in EUVI data.

In conclusion, our study supports the interpretation of stealth CMEs as a manifestation of normal solar eruptions, as opposed to belonging to a completely different class of solar phenomena. As events characterized by weak energetics, the magnetic structures associated to stealth CMEs seem to require a particular environment to successfully erupt, provided, in the case reported here, by a high-altitude null point. Future studies will investigate whether this is a general property of stealth CMEs.

\acknowledgments

JO thanks the STFC for support via funding given in her PHD studentship and would like to thank IAFE and the UCL Research Catalyst Awards sponsored by Santander Universities for support during her visit to IAFE.
LMG acknowledges support through a Royal Society University Research Fellowship. 
DML is grateful to the Science Technology and Facilities Council for the award of an Ernest Rutherford Fellowship (ST/R003246/1).
CHM and CMC acknowledge financial support from the Argentine grants PICT 2016-0221 (ANPCyT) and UBACyT 20020170100611BA. CHM is a member of the Carrera del Investigador Cient\'\i fico of the Consejo Nacional de Investigaciones Cient\'\i ficas y T\'ecnicas (CONICET). CMC is a fellow of CONICET.
GV acknowledges the support  from the European Union's Horizon 2020 research and innovation programme under grant agreement No 824135 and of the STFC grant number ST/T000317/1. The authors thank the anonymous referee whose comments helped to improve the paper.
SDO is a mission of NASA’s Living With a Star Program. STEREO is the third mission in NASA’s Solar Terrestrial Probes program. SOHO is a mission of international cooperation between ESA and NASA. The authors thank the SDO, STEREO, and SOHO teams for making their data publicly accessible.
We recognise the collaborative and open nature of knowledge creation and dissemination, under the control of the academic community as expressed by Camille Noûs at http://www.cogitamus.fr/indexen.html.
%\pc{This is a sentence that I have already put in my last published paper: 2020, A\&A, 639, A6. }

%\newpage
\bibliographystyle{agsm}
\bibliography{Main.bbl}

\end{document}